\documentclass[twocolumn,showpacs,amsmath,amssymb,pra,graphics,graphicx]{revtex4-2}
       \usepackage{graphicx}
        \usepackage{graphics}
\usepackage{bm}
\usepackage{dcolumn}
\usepackage{color}
\begin{document}

\title{Current distributions by moving vortices in superconductors}

\author{V. G. Kogan}
\email{kogan@ameslab.gov}
 \affiliation{Ames Laboratory--DOE, Ames, IA 50011, USA}
 \author{N. Nakagawa}
 \affiliation{Center for Nondestructive Evaluation, Iowa State University, Ames, IA 50011, USA}   

  \date{\today}
       
\begin{abstract}
We take account of normal currents that emerge when vortices move. Moving Abrikosov vortices  in the bulk and Pearl vortices in thin films are considered.
Velocity dependent distributions of both normal and persistent currents  are studied in the frame of time-dependent London equations. In thin films near the Pearl vortex core, these distributions  are intriguing in particular.
   \end{abstract}

\maketitle

\section{Introduction}
 
It has been shown recently   that 
moving vortex-like topological defects in, e.g., neutral superfluids or liquid crystals differ from their static versions \cite{leo}.   This is also the case in superconductors within the Time Dependent London theory (TDL) which takes into account normal currents, a necessary consequence of moving  vortex magnetic structure  \cite{TDL,KP-inter,KNak}. 
 
The London equations were employed for a long time to describe static or nearly static vortices out of their cores \cite{deGennes,Blatter}. The equations are linear, express the basic Meissner effect, and can be used at any temperature for problems where  the cores are irrelevant. As far as their current distributions are concerned, moving vortices are commonly considered the same as static    displaced as a whole.

 As shown in \cite{KNak}, the magnetic flux carried by moving vortex is equal to the flux quantum, but   is redistributed so that the part of it in front of the vortex is depleted whereas the part behind it is enhanced. This redistribution is caused by the normal currents resulting from the electric field induced by the moving non-uniform magnetic structure.  

In this paper,  analytic solutions are given for the field and current distributions   of Abrikosov vortices moving in the bulk.  It is shown that despite the anisotropic distribution of normal currents,   the ``topology" of supercurrents  in the vicinity of the vortex core remains close to the static case with nearly cylindrical  symmetry. This suggests that distortions of the vortex core shape in the bulk are weak. 
 
The case of   Pearl vortices  moving in thin films is quite different. For one, the sheet currents $\bm g$ in the film are related to the tangential field components $h_{x,y}$, rather than to $h_z$ of the vortex in the bulk. In other words, they are determined by the stray field in the free space out of the film and, as a result,   decay as a power law with the distance $r$ from the vortex core as compared to the exponential decay in the bulk. Besides, as $r\to 0$, the field $h_z\sim 1/r$, i.e. faster than $h_z\sim \ln (\lambda/r)$ in the bulk. Moreover, the sheet currents diverge as $1/r^2$ instead of bulk's $1/r$. For any in-plane anisotropy, the  supercurrents increase as $r\to 0$ leads to a situation in which the depairing value  is reached at different distances for different directions  that may lead to deviations of the vortex core shape from a circle of isotropic case. In particular, vortex motion is a source of such anisotropy, the subject of this work. We find the distribution of supercurrent values near the core of moving vortices is quite unconventional, see Figs.\,\ref{f10} and \ref{f11}. 

The time-dependent London equations are employed in this paper. The  equations are linear and contain the first derivative with respect to time, that makes them in a sense similar to the well-studied diffusion equation. As in diffusion case, we employ the Fourier transform in solving for fields and currents. To get results in real space, one has to calculate integrals over $k_x, k_y$ of the Fourier space from $-\infty$ to $\infty$, a heavy numerical procedure. We offer a way to transform the double integrals over $\bm k$ to single integrals from $0$ to $\infty$ which are easily evaluated numerically. \\ 
 
        In   time dependent situations,   the current   consists, in
general, of  normal and superconducting parts:
\begin{equation}
{\bm J}= \sigma {\bm E} -\frac{2e^2 |\Psi|^2}{mc}\, \left( {\bm
A}+\frac{\phi_0}{2\pi}{\bm
\nabla}\chi\right)  \,,\label{current}
\end{equation}
where ${\bm E}$ is the electric field, $\Psi$ is the order parameter, and $\chi$ is the phase.  

The conductivity  $\sigma$ approaches the normal state value  $\sigma_n$
when the temperature $T$ approaches $T_c$; in  s-wave
superconductors  it vanishes  fast with decreasing temperature along with the density of normal excitations. This is not the case for strong pair-breaking when   superconductivity becomes gapless while the density of states approaches the normal state value at all temperatures. Unfortunately, there is not much experimental information about the $T$ dependence of $\sigma$. Theoretically, this question is still debated, e.g.  Ref.\,\cite{Andreev} discusses possible enhancement of $\sigma$ due to inelastic scattering.

Within the London approach $|\Psi|$ is a constant $ \Psi_0 $ and Eq.\,(\ref{current})
becomes:
\begin{equation}
\frac{4\pi}{c}{\bm J}= \frac{4\pi\sigma}{c} {\bm E} -\frac{1}{\lambda^2}\,
\left( {\bm A}+\frac{\phi_0}{2\pi}{\bm
\nabla}\chi\right)  \,,\label{current1}
\end{equation}
where $\lambda^2=mc^2/8\pi e^2|\Psi_0|^2 $ is the London penetration depth.
Acting on this by curl one obtains:
\begin{equation}
{\bm h}- \lambda^2\nabla^2{\bm h} +\tau\,\frac{\partial {\bm h}}{\partial
t}= \phi_0 {\bm z}\sum_{\nu}\delta({\bm r}-{\bm r_\nu})\,,\label{TDL}
\end{equation}
where $\phi_0$ is the flux quantum, ${\bm r_\nu}(t) $ is the position of the $\nu$-th vortex, $\bm z$ is the direction of vortices, and the relaxation time 
\begin{equation}
\tau=  4\pi\sigma\lambda^2/c^2 \,.
\label{tau}
\end{equation}
Equation (\ref{TDL}) can be considered as a general form of the time
dependent London equation. 
 
Within the general approach to slow relaxation processes one has
 \begin{eqnarray}
 \gamma \frac{\partial {\bm h}}{\partial t}=-\frac{\delta  F }{\delta {\bm h}}\,,  
  \label{B1}
\end{eqnarray}
where $F={\cal F}/V=\int d^2{\bm r}\left(h^2+\lambda^2({\rm curl} {\bm h})^2\right)/8\pi V$ is the London free energy density (sum of magnetic and kinetic parts). Comparing this with Eq.\,(\ref{TDL})  one has $\gamma=\sigma\lambda^2/c^2=4\pi\tau$. In fact, the time-dependent GL equations can be obtained in a similar  manner \cite{Kopnin-Gor'kov}.

Like in the static London approach, the time dependent version of London equations (\ref{TDL}) is valid only outside vortex cores.  As such it may give useful results for materials with large Ginzburg-Landau parameter $\kappa$ in fields away of the upper critical field $H_{c2}$. On the other hand, Eq.\,(\ref{TDL}) is a useful, albeit approximate, tool for low temperatures where the Ginzburg-Landau theory does not work and the microscopic theory is forbiddingly complex.

\section{ Moving Abrikosov vortex  in the bulk}

The field distribution of this case has been evaluated numerically in Ref.\,\cite{TDL}.  Here, we provide this distribution in a closed analytic form. 

 The magnetic field $\bm h$ has one component $h_z$, so we can omit the subscript $z$. Taking the Fourier transform of Eq.\,(\ref{TDL}) and solving the differential equation for $h_{\bm k}(t)$ we obtain at $t=0$:
 \begin{equation}
   h({\bm r}) = \frac{\phi_0}{4\pi^2\lambda^2}   \int  \frac{d^2{\bm k}\,e^{ i{\bm k} {\bm r} } }{1+ k^2-i
k_x s}\,,\qquad s=\frac{v\tau}{\lambda}\,,
\label{h}
\end{equation}
where $\lambda$ is chosen as a unit length in writing the dimensionless integral. To evaluate this double integral,
 we use the identity
\begin{eqnarray}
 ( 1+  k^2-i  k_x s)^{-1}= \int_0^\infty e^{-u( 1+  k^2-i  k_x s)} du\,,
\label{identa} 
\end{eqnarray}
so that
\begin{eqnarray}
&&\frac{4\pi^2\lambda^2}{\phi_0}  h (\bm r)   = \int_0^\infty du\,e^{-u}  \int  d^2\bm k\, e^{i\bm k\cdot\bm r-u (k^2-ik_xs) } \nonumber\\
&&=  \int_0^\infty du\,e^{-u}  \int  d^2\bm k\, e^{i\bm k\cdot\bm \rho-u k^2  },\,\,\, \bm \rho=(x+us,y) .\qquad
\label{hz(r)1a} 
\end{eqnarray}
Integrals over $k_x,k_y $ are Fourier transforms of Gaussians: 
\begin{eqnarray}
  \int_{-\infty}^\infty   dk_x\, e^{i k_x \rho_x-u k_x^2  } \int_{-\infty}^\infty   dk_y\, e^{i k_y y-u k_y^2  }=\frac{\pi}{u}e^{-\rho^2/4u}.\qquad
\label{int} 
\end{eqnarray}
 Hence, we have
 \begin{eqnarray}
  h (\bm r)   
  &=&\frac{\phi_0}{4\pi \lambda^2} \int_0^\infty \frac{du\,e^{-u}}{u}\,\exp\left[-\frac{(x+us)^2+y^2}{4u}\right] \nonumber\\
  &=& \frac{\phi_0}{2\pi\lambda^2} e^{-sx/2\lambda} K_0\left(\frac{r}{2\lambda}\sqrt{4+s^2}\right) \qquad
\label{hz(r)1b} 
\end{eqnarray}
where $K_0$ is the Modified Bessel function and the last line is written in common units 
 \cite{remark}.
Note that for the vortex at rest $s=0$ and we get the standard result $h=(\phi_0/2\pi\lambda^2) K_0(r/\lambda)$ \cite{deGennes}.  

The current distribution follows:
 \begin{eqnarray}
  \frac{8\pi^2\lambda^3}{c\phi_0}  j_x  &=& -  \frac{y\sqrt{4+s^2}}{2r} e^{-sx/2}K_1\left(\frac{r}{2 }\sqrt{4+s^2}\right),   \label{Jx}\\
    \frac{8\pi^2\lambda^3}{c\phi_0}  j_y  &=&  e^{-sx/2}\Big[ \frac{ s}{2 } K_0\left(\frac{r}{2 }\sqrt{4+s^2}\right) \nonumber\\
  &+& \frac{x\sqrt{4+s^2}}{2r} K_1\left(\frac{r}{2 }\sqrt{4+s^2}\right)  \Big ],  
    \label{Jy} 
\end{eqnarray}
where $\lambda$ is used as a unit length on the right-hand sides.

The current $\bm j$ here is obtained from the field $\bm h$, so that it is the {\it total}, superconducting and  normal,
   $\bm j=\bm j_s +\bm j_n$. It is of interest to have also $ \bm j_s$ and $\bm j_n$ separately. To this end, we note that $\bm j_n=\sigma\bm E$, so that the stream lines of $\bm j_n$ coincide with those for $\bm E$. Hence, one takes the Fourier transform of the  field $\bm h$ from  Eq.\,(\ref{h}) and obtains the electric field with the help of Maxwell equations $i({\bm k}\times {\bm E}_{\bm k})_z=- \partial_t h_{z{\bm k}}/c$
and ${\bm k}\cdot{\bm E}_{\bm k}=0$:
\begin{eqnarray}
\frac{c}{\phi_0v}  E_{x{\bm k}} &=&  -\frac{k_xk_y }{k^2(1+k^2 -ik_x  s )}  \,,\label{Exk} \\
\frac{c}{\phi_0v}  E_{y{\bm k}}&=&   \frac{k_x^2 }{k^2(1+k^2 -ik_x  s)}  \,.\label{Eyk}
\end{eqnarray}
($\lambda$ is used as the unit length).

The  field $\bm E(\bm r)$ in real space can be obtained in the same manner as was done   for $h(\bm r)$. The results are:
 \begin{eqnarray}
E_x &=&\frac{\phi_0v}{2\pi c\lambda^2}\int_0^\infty du\,e^{-u}\frac{2y(x+us)}{[(x+us)^2 +y^2]^2 } \,,\label{Ex(r)}\\
 E_y &=&\frac{\phi_0v}{2\pi c\lambda^2}\int_0^\infty du\,e^{-u}\frac{ y^2-(x+us)^2}{[(x+us)^2 +y^2]^2 } \,.
 \label{Ey(r)}
\end{eqnarray}

 The stream lines of   $\bm E$  satisfy $0=(d\bm \ell \times \bm E)_z=dx E_y-dy E_x$ where $d\bm \ell=(dx,dy)$  is a line element. Introduce now a scalar ``stream function"   $G(x,y)$ such that
 \begin{eqnarray}
  E_y=\frac{\partial G}{\partial x},\qquad  E_x=-\frac{\partial G}{\partial y}\,,
\label{G1a}
 \end{eqnarray}
i.e., the stream lines   are given by $G(x,y)=\,\,$const. 
One can check by direct differentiation that 
  \begin{eqnarray}
G(\bm r) =\frac{\phi_0v}{2\pi c\lambda^2 }\int_0^\infty \frac{du\,e^{-u}(x+us)}{(x+us)^2 +y^2 }    
\label{G4a}
\end{eqnarray}
generates $\bm E$ of Eqs.\,(\ref{Ex(r)}) and (\ref{Ey(r)}). 

\begin{figure}[h ]
\includegraphics[width=7.5cm] {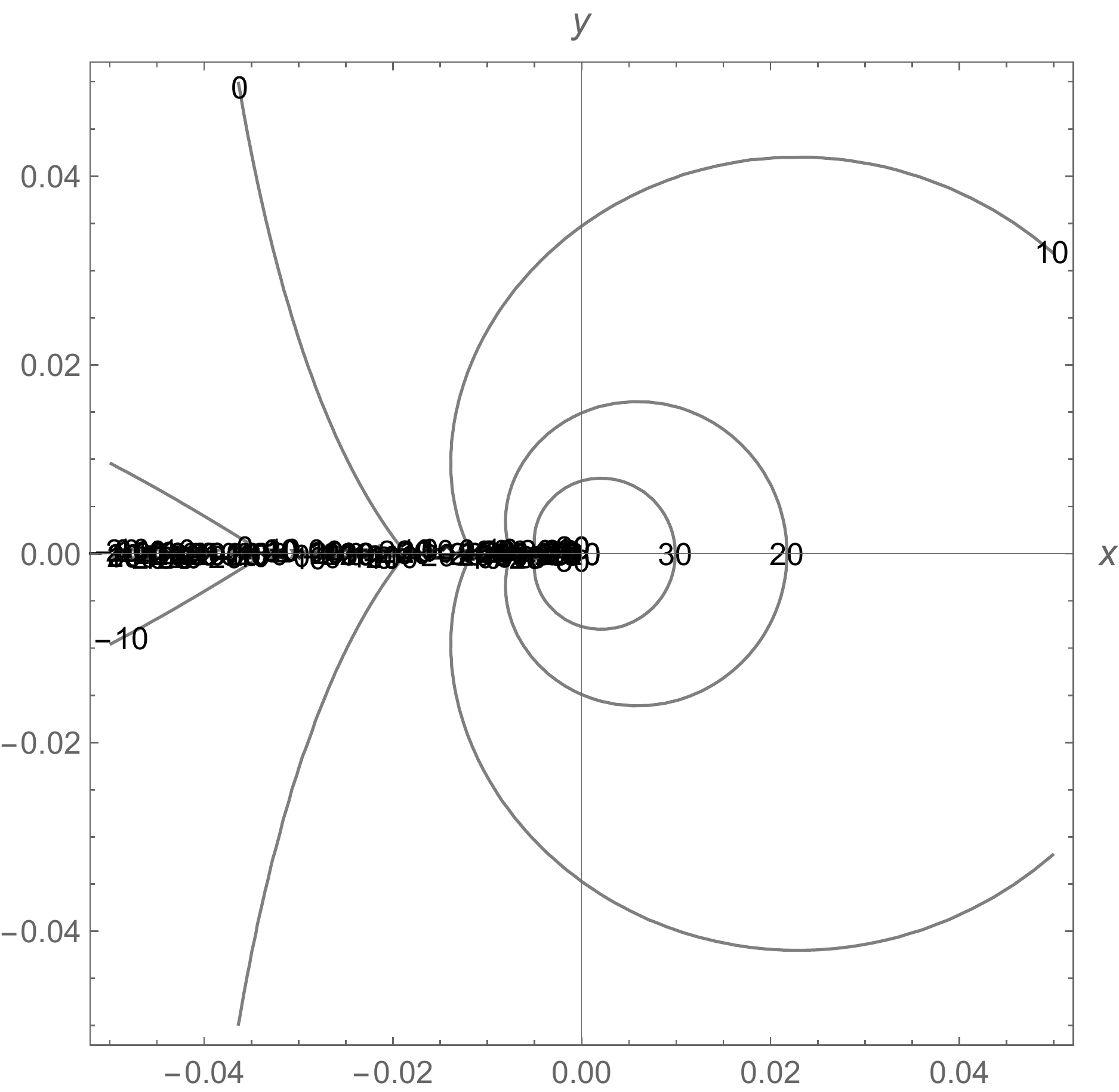}
\caption{The stream lines of the normal current near the moving vortex for $s=0.05$.  Positive contour numbers correspond to the clockwise direction of $\bm j_n$ whereas negative to the counterclockwise;  $x,y$  are in units of $\lambda$.  The cut at $x<0$ of the $x $ axis is traced to Eq.\,(\ref{Ey(r)}) for $E_y$ where at $y=0$ and negative $x$ the integral is not defined. $x,y$ are in units of $\lambda$.}
\label{f1}
\end{figure}

Fig.\,\ref{f1}  shows  the stream lines of normal currents for $s=0.05$.
Since the field $\bm h$ is directed out of the figure plane, the normal currents   cause  its reduction in front of the moving vortex and enhancement behind.
 
The total magnetic flux carried by a vortex is still $\phi_0$, so that appearance of normal currents should change the distribution of supercurrents as well: 
 \begin{eqnarray}
\bm j_s=\bm j-\bm j_n = \frac{c\phi_0 }{8\pi^2\lambda^3}\left(\hat{\bm j} -\frac{s}{2\pi}\hat{\bm  E} \right) 
\label{Js}
\end{eqnarray}
where the under-hat quantities are dimensionless RHS's in expressions  (\ref{Jx}), (\ref{Jy}) for $\bm J$ and integrals in Eqs.\,(\ref{Ex(r)}), (\ref{Ey(r)}) for $\bm E$. 

Of a particular interest is the distribution of the {\it values} $|\bm j_s|=\sqrt{j_{sx}^2+j_{sy}^2}$, because   $|\bm j_s|$ cannot exceed the depairing value, thus defining qualitatively the ``core boundary".
   \begin{figure}[h ]
\includegraphics[width=7.5cm] {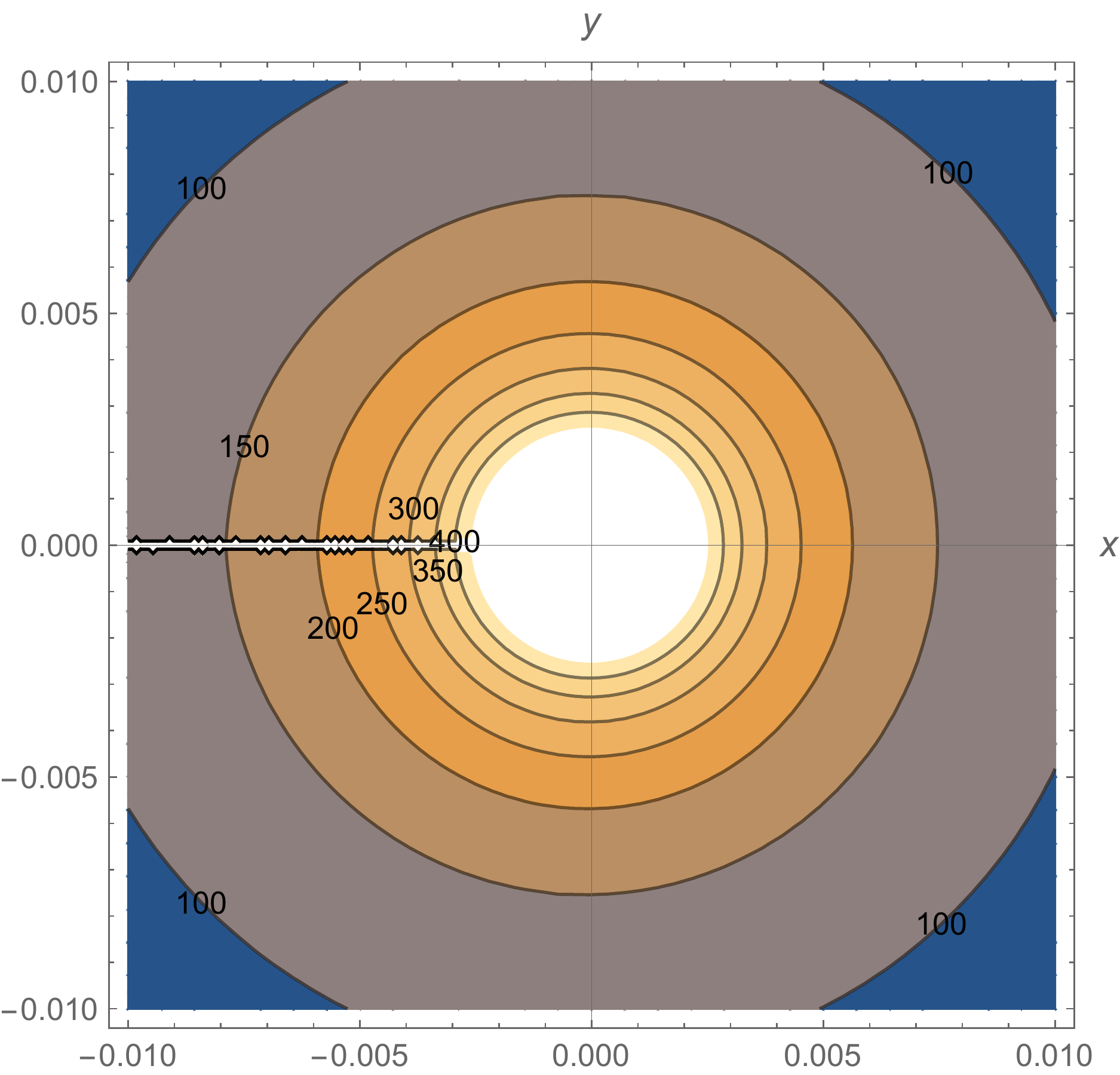}
\caption{(Color online) Contours of constant values of the supercurrent $|j_s|$ near the moving vortex for $s=0.05$;  $x,y$  are in units of $\lambda$. }
\label{f2}
\end{figure}

Fig.\,\ref{f2} shows contours of constant $|j_s|$ for $s=0.05$ in the vicinity of the core. 
Hence, the moving vortex core in this case should be close to a circle; the anisotropy of $j_s$ is still seen, e.g., in the contour marked by 100, which is only slightly differs from a circle. In other words, despite the presence of normal currents lacking any resemblance of cylindrical symmetry, Fig.\,\ref{f1}, the core shape of Abrikosov vortex  is hardly affected by the vortex motion. We have checked that for a faster motion with $s=2$, to find $j_s(x,y)=\,\,$const  near the singularity is close to a circle. This implies that for a moving Abrikosov vortex in the bulk the normal currents near the core are small and their effects on persistent currents are weak.
As is shown below, the situation in thin films is drastically different.
  
\section{Thin films}

We start this section with  a few known general results and then apply them to moving Pearl vortices.
 Let the film of thickness $d$ be in the $xy$ plane. Integration of Eq.\,(\ref{TDL}) over the film  thickness gives for the $z$ component of the field at the film for a Pearl vortex moving with velocity $\bm v$:
\begin{eqnarray}
\frac{2\pi\Lambda}{c}{\rm curl}_z {\bm g} + h_z  +\tau\frac{\partial h_z}{\partial t}=\phi_0 \delta(\bm r -\bm  vt)    .
\label{2D London}
\end{eqnarray}
Here,  $\bm g$ is the sheet current density related to the tangential field components at the upper film face by  $2\pi\bm g/c=\hat{\bm z}\times \bm h$; $\Lambda=2\lambda^2/d$ is the Pearl length, and $\tau=4\pi\sigma\lambda^2/c^2$. With the help of div$\bm h=0$ this equation is transformed to:
\begin{eqnarray}
h_z -\Lambda \frac{\partial h_z}{\partial z}   +\tau\frac{\partial h_z}{\partial t}=\phi_0 \delta(\bm r -\bm  vt) .
\label{hz-eq}   
\end{eqnarray}
 
A  large contribution to the energy of a vortex in a thin film comes from  stray fields \cite{Pearl}. The problem of a vortex in a thin film is, in fact, reduced to that of the field distribution in free space subject to the boundary condition supplied by solutions of Eq.\,(\ref{2D London}) at the film surface. Since outside the film curl$\bm h=\,\,\,$div$\bm h=0$ (see remark  \cite{remark2}), one can introduce a scalar potential for the {\it outside} field in the upper half-space:
 \begin{eqnarray}
\bm h   =\bm \nabla \varphi,\qquad \nabla^2\varphi=0   \,.
\label{define _phi} 
\end{eqnarray}
The general form of the potential satisfying Laplace equation that vanishes at $z\to\infty$   is 
 \begin{eqnarray}
\varphi (\bm r, z)   =\int \frac{d^2\bm k}{4\pi^2} \varphi(\bm k) e^{i\bm k\cdot\bm r-kz}\,.
\label{gen_sol} 
\end{eqnarray}
Here, $\bm k=(k_x,k_y)$, $\bm r=( x, y)$, and  $ \varphi(\bm k)$  is the two-dimensional Fourier transform of $ \varphi(\bm r, z=0)$. In the lower half-space one has to replace $z\to -z$ in Eq.\,(\ref{gen_sol}). 

As is done in \cite{TDL}, one applies the 2D Fourier transform to Eq.\,(\ref{hz-eq}) to obtain a linear differential equation for $h_{z\bm k}(t)$. The solution is
\begin{eqnarray}
h_{z\bm k}   =-k\varphi_{\bm k}= \frac{\phi_0 e^{-i\bm k\cdot\bm v t}}{ 1+\Lambda k-i \bm k\cdot\bm v \tau }  \, 
\label{hz(k)} 
\end{eqnarray}
and
\begin{eqnarray}
\varphi_{\bm k}   =-\frac{\phi_0 e^{-i\bm k\cdot\bm v t}}{k(1+\Lambda k-i \bm k\cdot\bm v \tau) }  \,.
\label{phi(k)} 
\end{eqnarray}
In fact, this gives distributions of all field components outside the film, its surface included. 
 
We are interested in the vortex motion  with constant velocity $\bm v=v\hat{\bm x}$, so that we can evaluate this field in real space for the vortex at the origin at $t=0$:
\begin{eqnarray}
h_{z}(\bm r)   = \frac{\phi_0}{4\pi^2} \int \frac{d^2\bm k \,e^{i\bm k\cdot\bm r}}{ 1+\Lambda k-i  k_x v\tau }\,.
\label{hz(r)} 
\end{eqnarray}
 It is convenient in the following to use Pearl $\Lambda$ as the unit length and measure the field in units $\phi_0/4\pi^2\Lambda^2$: 
 \begin{eqnarray}
h_{z}(\bm r)   =  \int \frac{d^2\bm k \,e^{i\bm k\cdot\bm r}}{ 1+ k-i  k_x s }\,, \label{hz(r)a} 
\end{eqnarray}
 we left the same  notations for $\bm h_z$ and $\bm k$ in new units; when needed, we   indicate formulas  written in common units. 
 The parameter 
  \begin{eqnarray}
 s=\frac{v\tau}{\Lambda}=2\pi \frac{v\sigma d}{c^2}\, ,
\label{s} 
\end{eqnarray}
so that $s\ll 1$ for most practical situations. However, in recent experiments the velocities 
 $ 1.5\times 10^6\,$cm/s were recorded, for which $s$ might reach $0.1$ which may still increase in future experiments \cite{Eli,Denis}. Besides, vortices exist not only in low temperature laboratories, but e.g. in neutron stars about their motion we know little. Hence, in the derivations below we consider  arbitrary values of $s$.

After  applying the same formal procedure as  for 3D Abrikosov vortex  one obtains \cite{KNak}:
 \begin{eqnarray}
  h_{z}(\bm r)   =2\pi   \int_0^\infty du \frac{u\,e^{-u} }{(\rho^2+u^2)^{3/2}}  
\label{hz(r)d} 
\end{eqnarray}
with $\rho^2=(x+us)^2+y^2$.  Hence, we succeeded in reducing the double integral (\ref{hz(r)a}) to a single integral over $u$ which is readily evaluated. 

\begin{figure}[t]
\includegraphics[width=7.5cm] {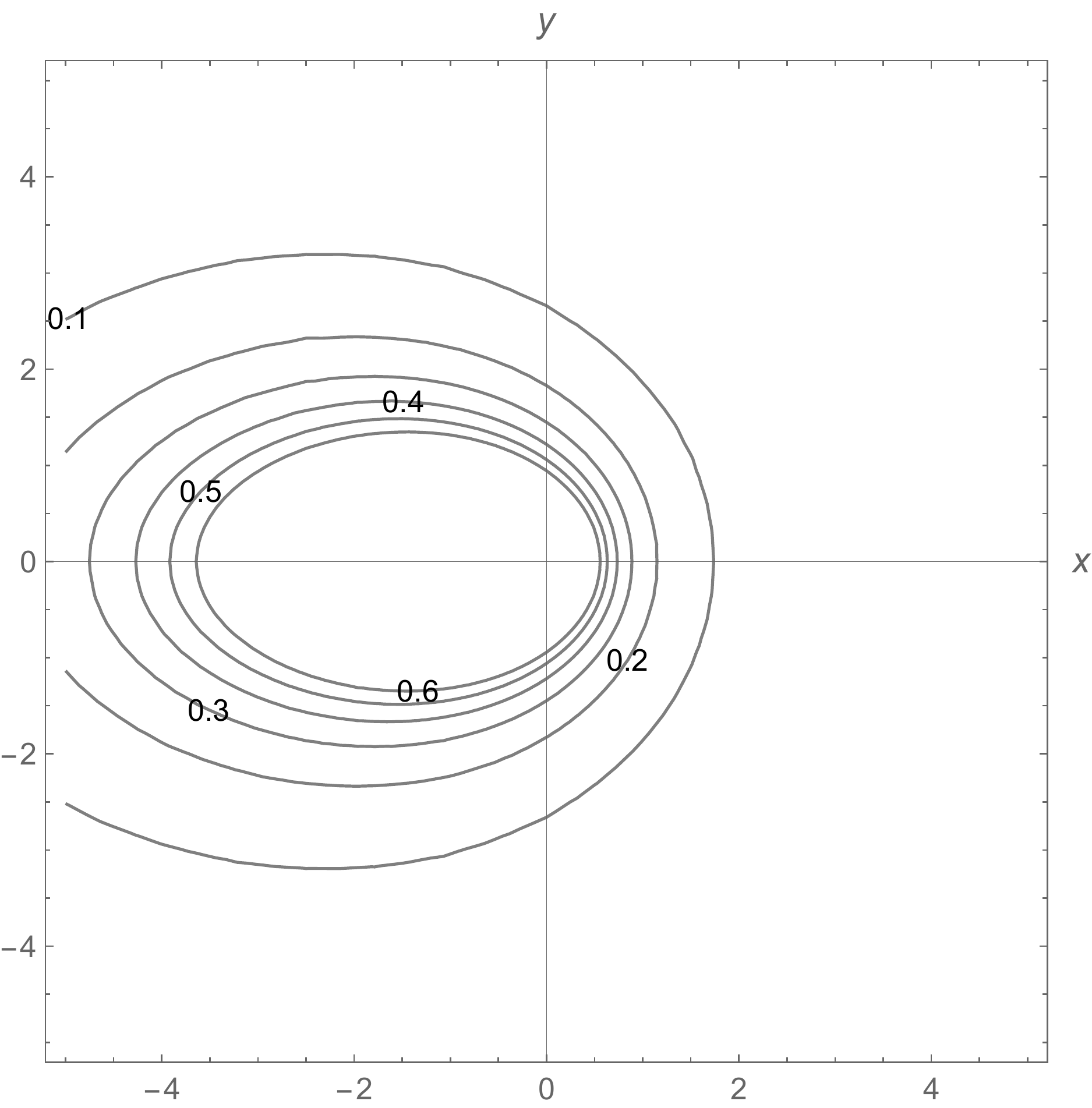}
\caption{ Contours of  $h_z(x,y)=\,\,$const ($h_z$ is in units  $\phi_0/4\pi^2\Lambda^2$ and $x,y$ in units of $\Lambda$)  for $s=2$.  $x,y$ are in units of $\Lambda$.}
\label{f3}
\end{figure}

The results are shown in Fig.\,\ref{f3}.  
The field distribution is not symmetric relative to the singularity position: the field in front of the moving vortex is suppressed relative to the symmetric distribution of the vortex at rest, whereas behind the vortex it  is enhanced  \cite{KNak}. 

Integrating by parts we obtain from Eq.\,(\ref{hz(r)d}):
    \begin{eqnarray}
 h_{z}   =  2\pi  \left[ \frac{1}{r}-  
\int_0^\infty   \frac{du\,e^{-u} }{\sqrt{\rho^2+u^2}} \left(1+\frac{s(x+su)}{\rho^2+u^2} \right)\right] .\qquad
\label{hz(r)2} 
\end{eqnarray}
For the Pearl vortex at rest $s=0$, $ \rho=  r$, and the  known result for a vortex at rest follows, see e.g. Ref.\,\cite{jjf}. In particular, the last form of $h_z(x,y)$ shows that as $r\to 0$ the leading term of this field  diverges as $1/r$, i.e., faster than $|\ln r|$ of the Abrikosov vortex in the bulk \cite{jjf,Tafuri-Kirtley}.

For the most realistic slow motion with $s\ll 1$, one can get analytic approximation for the field distribution. To this end, go back to Eq.\,(\ref{hz(r)d})  and expand the integrand in powers of small $s$ up to ${\cal O}(s)$:
    \begin{eqnarray}
 h_{z}   =  2\pi  \left( 
\int_0^\infty   \frac{du\,u\,e^{-u} }{(r^2+u^2)^{3/2}} +3sx\int_0^\infty   \frac{du\,u^2 e^{-u} }{(r^2+u^2)^{5/2}} 
 \right)  .\qquad
\label{hz(r)2B} 
\end{eqnarray}
The first term gives the field  of the vortex at rest. 
The second term is:
\begin{eqnarray}
-s\, x\left[ \frac{2}{r}+  \frac{\pi}{2} ( Y_0 -\bm H_0)_r +   \frac{\pi}{2r} ( Y_1 -\bm H_1)_r
 \right ]      \,,
\label{correction} 
\end{eqnarray}
where $  Y_{0,1}$ and $\bm H_{0,1}$ are Bessel and Struve functions of  argument $r$.

\section{Current distribution}

As mentioned above, the sheet current is related to the tangential field components by
\begin{eqnarray}
\frac{2 \pi}{c}g_x = -h_y\,,\qquad \frac{2 \pi}{c}g_y =  h_x\,.
\label{current1} 
\end{eqnarray}
In 2D Fourier space,  tangential fields are  $h_{x\bm k}=ik_x\varphi_{\bm k}$ and $h_{y\bm k}=ik_y\varphi_{\bm k}$.  
The potential at $t=0$ and $z=+0$  is given (in common units) by 
 \begin{eqnarray}  
 \varphi_{\bm k} =-  \frac{\phi_0 }{k( 1+  \Lambda k-i  k_x v\tau)}.
\label{energy} 
\end{eqnarray}
Then, we have:
\begin{eqnarray}
 g_x(\bm r,s) = - i\frac{c\phi_0}{2\pi\Lambda^2} I_x\,,\quad I_x =\int\frac{d\bm k\, k_ye^{i\bm k\bm r}}{k(1+k-ik_x s)} , \qquad   \label{gx}\\
  g_y(\bm r,s) =  i\frac{c\phi_0}{2\pi\Lambda^2} I_y\,,\quad I_y =\int\frac{d\bm k\, k_xe^{i\bm k\bm r}}{k(1+k-ik_x s)}. \qquad
 \label{gy} 
\end{eqnarray}
To evaluate  dimensionless integrals $I_{x,y}$ we make use of the identity  (\ref{identa}) in which $k^2\to k$:
\begin{eqnarray}
  I_x &=&\int\frac{d\bm k\, k_ye^{i\bm k\bm r}}{k}\int_0^\infty du\, e^{-u(1+k-ik_x s)}\nonumber   \\
  &=&\int_0^\infty du\,e^{-u} \int \frac{d\bm k\, k_y}{k}e^{i\bm k \bm\rho -uk}
   \label{Ix}
\end{eqnarray}
To evaluate here the integral over $\bm k$, we make use of the Coulomb Green's function which can be manipulated to the form \cite{KNak}:
\begin{eqnarray}
  \frac{1}{\sqrt{x^2+y^2+z^2}}   =\frac{1}{2\pi}\int\frac{d^2\bm k}{ k} e^{i\bm k\bm r-kz}  . \qquad   
  \label{ident1}
\end{eqnarray}
Replace now $\bm r \to \bm\rho=(x+us,y)$, $z\to u$:
\begin{eqnarray}
  \frac{1}{\sqrt{(x+us)^2+y^2+u^2}}   =\frac{1}{2\pi}\int\frac{d\bm k}{ k} e^{i\bm k\bm \rho-ku}   \qquad   
  \label{ident2}
\end{eqnarray}
  \begin{figure}[b]
\includegraphics[width=7.5cm] {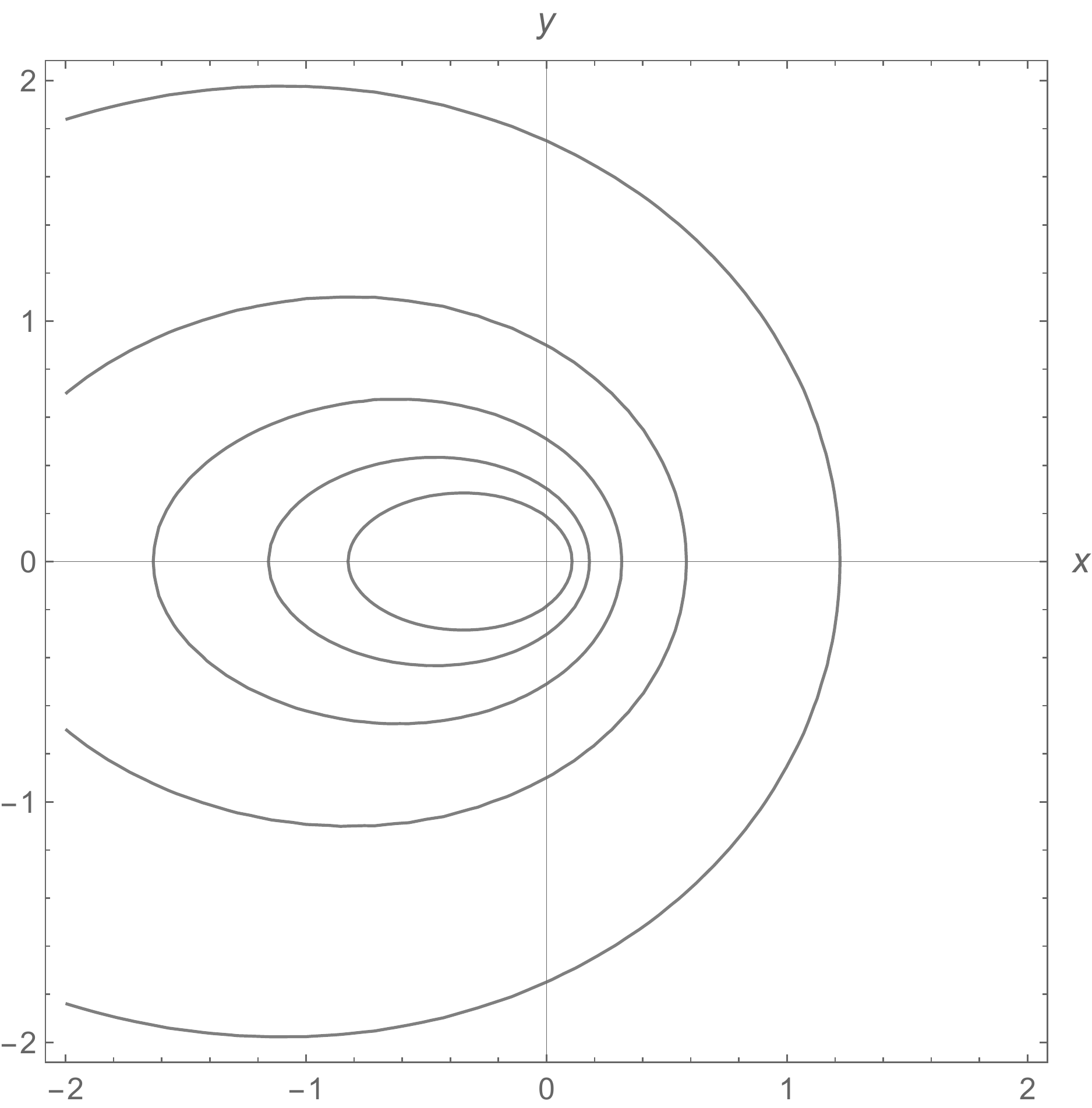}
\caption{  Contours of constant potential $\varphi(x,y)$ and, what is the same, the stream lines of the current $\bm g$ for a fast moving vortex, $s=2$.  $x$ and $y$ are in units of $\Lambda$. }
\label{f4}
\end{figure}
and apply $\partial_y$   to get   the integral over $\bm k$ in Eq.\,(\ref{Ix}):
 \begin{eqnarray}
 - \frac{y}{ (\rho^2+u^2)^{3/2}}  =\frac{i}{2\pi}\int\frac{d\bm k\,k_y}{ k} e^{i\bm k\bm \rho-ku}  . \qquad   
  \label{ident3}
\end{eqnarray}
Hence, we obtain
\begin{eqnarray}
\frac{\Lambda^2}{c\phi_0}\, g_x= y  \int_0^\infty\frac{du\,e^{-u}}{(\rho^2+u^2)^{3/2}} , \qquad   
  \label{gx1}
\end{eqnarray}
and 
\begin{eqnarray}
\frac{\Lambda^2}{c\phi_0}\, g_y=  - \int_0^\infty\frac{du\,e^{-u}(x+us)}{(\rho^2+u^2)^{3/2}}   . \qquad   
  \label{gy1}
\end{eqnarray}

It is easy to see that  stream lines of the total current  coincide with  contours of $\varphi(x,y)=\,\,$const. These are shown in Fig.\,\ref{f4}, so that for a fast motion the current distribution differs substantially from the static case. 
 
It is worth noting that the currents  in Eqs.\,(\ref{gx1}) and (\ref{gy1})  are in fact {\it total}, i.e. the sum of  persistent and  normal currents, $\bm g=\bm g_s+\bm g_n$.  It is of interest to separate these contributions, because  the size and shape of  vortex cores are related to the value of persistent currents only. As was done for the Abrikosov vortex, we calculate first the normal currents   $ \bm g_n=\sigma\bm Ed$. 

\subsection{Normal currents}

  \begin{figure}[h ]
\includegraphics[width=7.5cm] {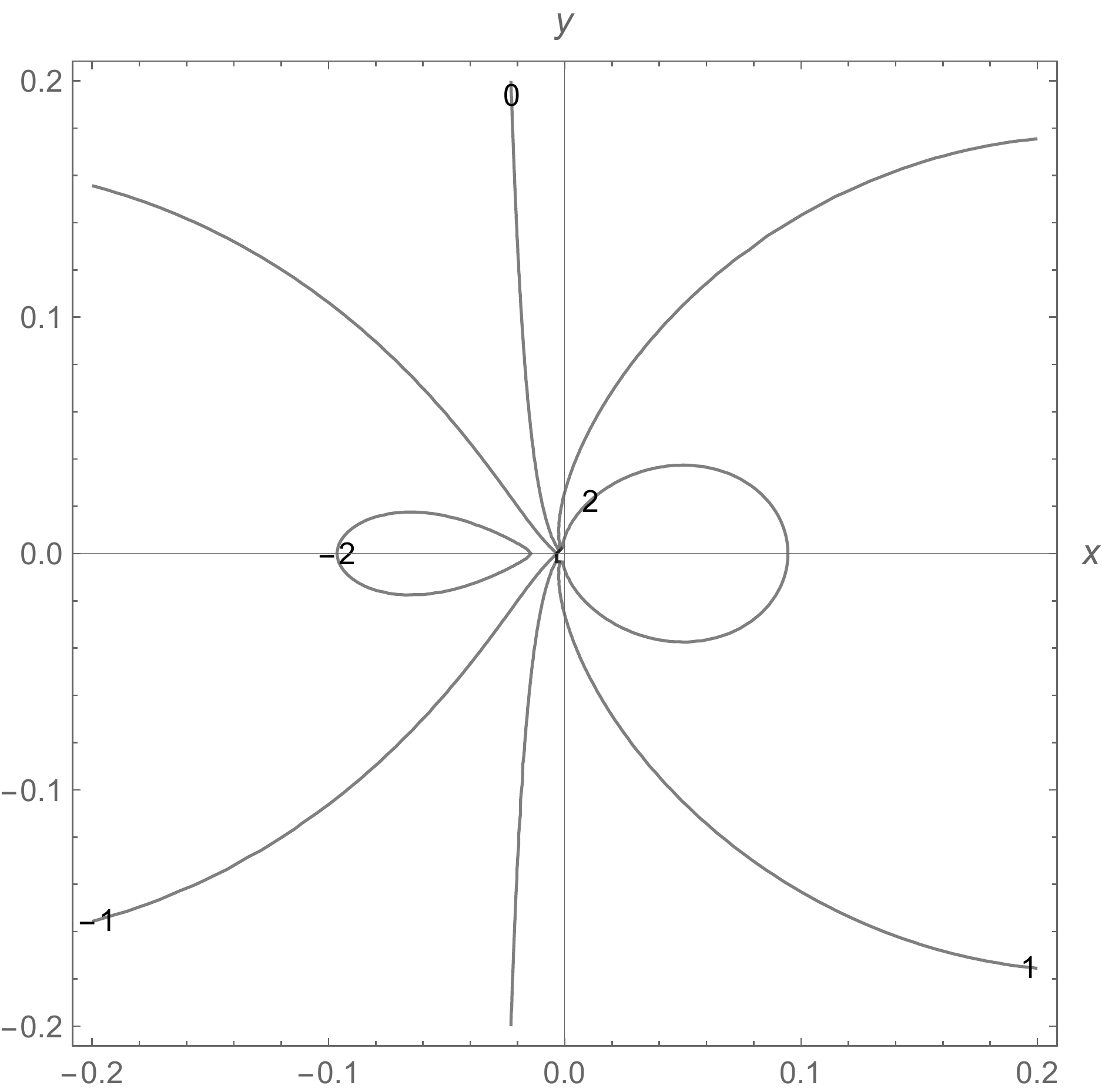}
\caption{ Stream lines of the normal current, i.e., the lines of $S(x,y)=\,\,$const, for a  slow motion with $s=0.05$.  $x$ and $y$ are units of $\Lambda$. The positive values of $S$ correspond to clockwise direction of $\bm g_n$ and negative for counterclockwise.}
\label{f5}
\end{figure}

To this end, one takes the magnetic field of moving vortex, Eq.\,(\ref{hz(k)}), and obtains the electric field from  Maxwell equations $i({\bm k}\times {\bm E}_{\bm k})_z=- \partial_t h_{z{\bm k}}/c$
and ${\bm k}\cdot{\bm E}_{\bm k}=0$:
\begin{eqnarray}
 E_{x{\bm k}} &=&  -\frac{\phi_0v}{c} \frac{k_xk_y }{k^2(
1+k -ik_x  s )}  \,,\label{exk} \\
E_{y{\bm k}}&=& \frac{\phi_0v}{c} \frac{k_x^2 }{k^2(
1+k -ik_x  s)}  \label{eyk}
\end{eqnarray}
($\Lambda$ is used as the unit length).

The stream lines of the normal current $\bm g_n$ coincide with those for $\bm E$ which satisfy $0=(d\bm \ell \times \bm E)_z=dx E_y-dy E_x$. Introducing   $S(x,y)$ such that
 \begin{eqnarray}
  E_y=\frac{\partial S}{\partial x},\qquad  E_x=-\frac{\partial S}{\partial y}\,,
\label{G1}
 \end{eqnarray}
we see that the stream lines of the vector field $\bm E$ are given by $S(x,y)=\,\,$const. Using Eqs.\,(\ref{exk})  or (\ref{eyk}) we  obtain in Fourier space
\begin{eqnarray}
S_{\bm k} = \frac{k_x}{ik^2 (1+k-ik_x s)}   
\label{G2}
\end{eqnarray}
and 
 \begin{eqnarray}
S(\bm r) = \int \frac{d\bm k\,k_xe^{i\bm k\bm r -kz}}{k^2 (1+k-ik_x s)}   
\label{G3}
\end{eqnarray}
where  constant pre-factors are omitted. Making use of  identity (\ref{ident2}) we arrive at
 \begin{eqnarray}
S(\bm r) &=& \int_0^\infty du\,e^{-u} \frac{x+us}{{\cal R}({\cal R}+u)}  \,,
\label{G4}\\
{\cal R}^2&=&(x+us)^2+y^2+u^2=\rho^2+u^2\,.
\label{calR} 
\end{eqnarray}
The electric field is now obtained with the help of Eq.\,(\ref{G1}):
\begin{eqnarray}
&&\frac{2\pi \Lambda^2c}{\phi_0v} E_x(\bm r)= - \int_0^\infty du\,e^{-u}  \frac{y(x+us) (u+2{\cal R}) }{
{\cal R}^3(u+{\cal R})^2} \,, \qquad
\label{Ex}\\
&&\frac{2\pi \Lambda^2c}{\phi_0v} E_y(\bm r) \nonumber\\
&&= \int_0^\infty du\,e^{-u}   \frac{{\cal R}^2 (u+{\cal R})-(x+us)^2 (u+2{\cal R})}{{\cal R}^3(u+{\cal R})^2} .\qquad   
\label{Ey}
 \end{eqnarray}
 
   \begin{figure}[h ]
\includegraphics[width=7.5cm] {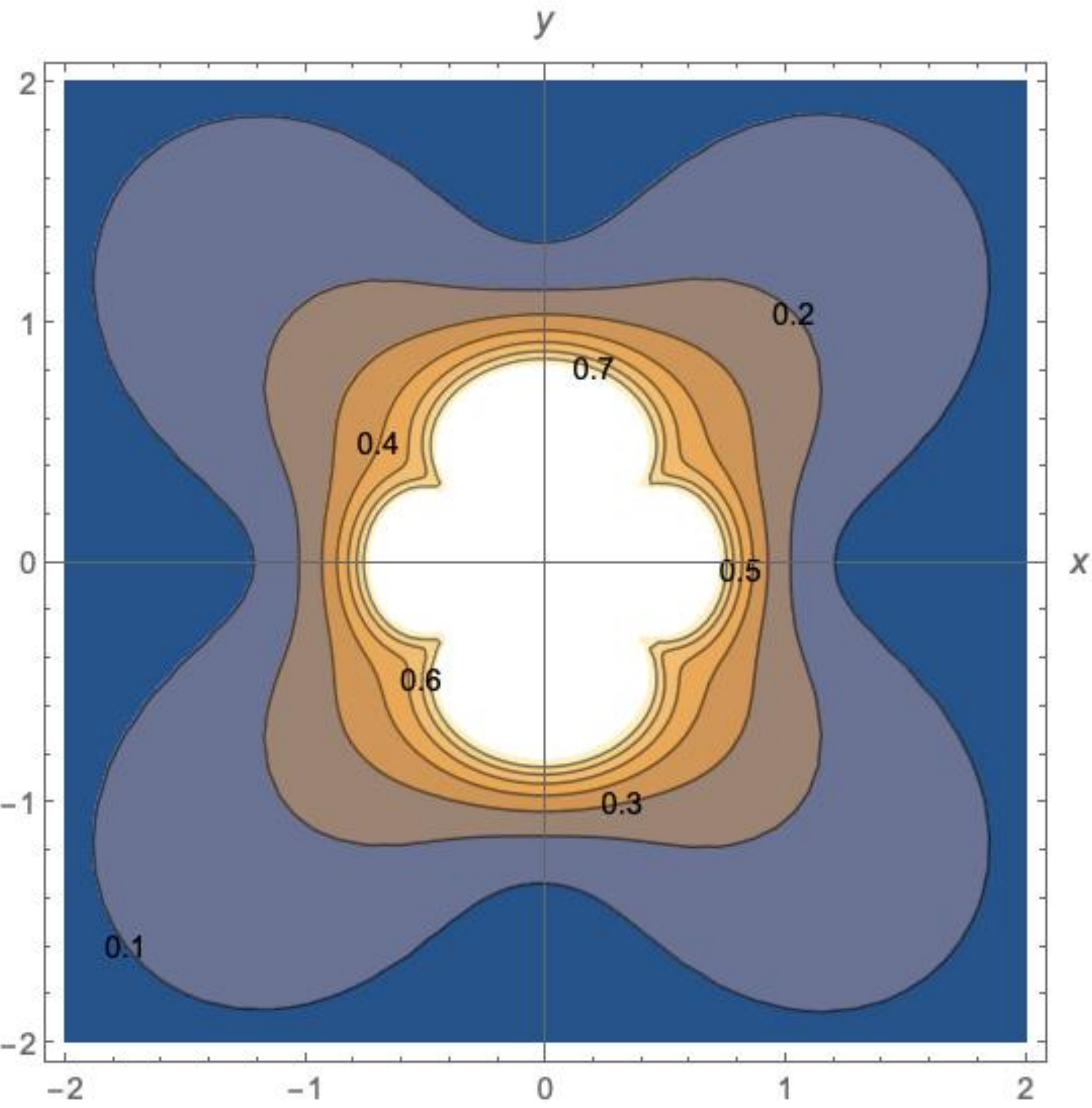}
\caption{(Color online)   Contours of constant normal current  {\it values} $\sqrt{g_{nx}^2 +g_{ny}^2 }$ for a  moving vortex with $s=0.05$.  $x$ and $y$ are in units of $\Lambda$, so that one can say this figure represents the distribution at large distances. }
\label{f6}
\end{figure}


  \begin{figure}[h ]
\includegraphics[width=7.5cm] {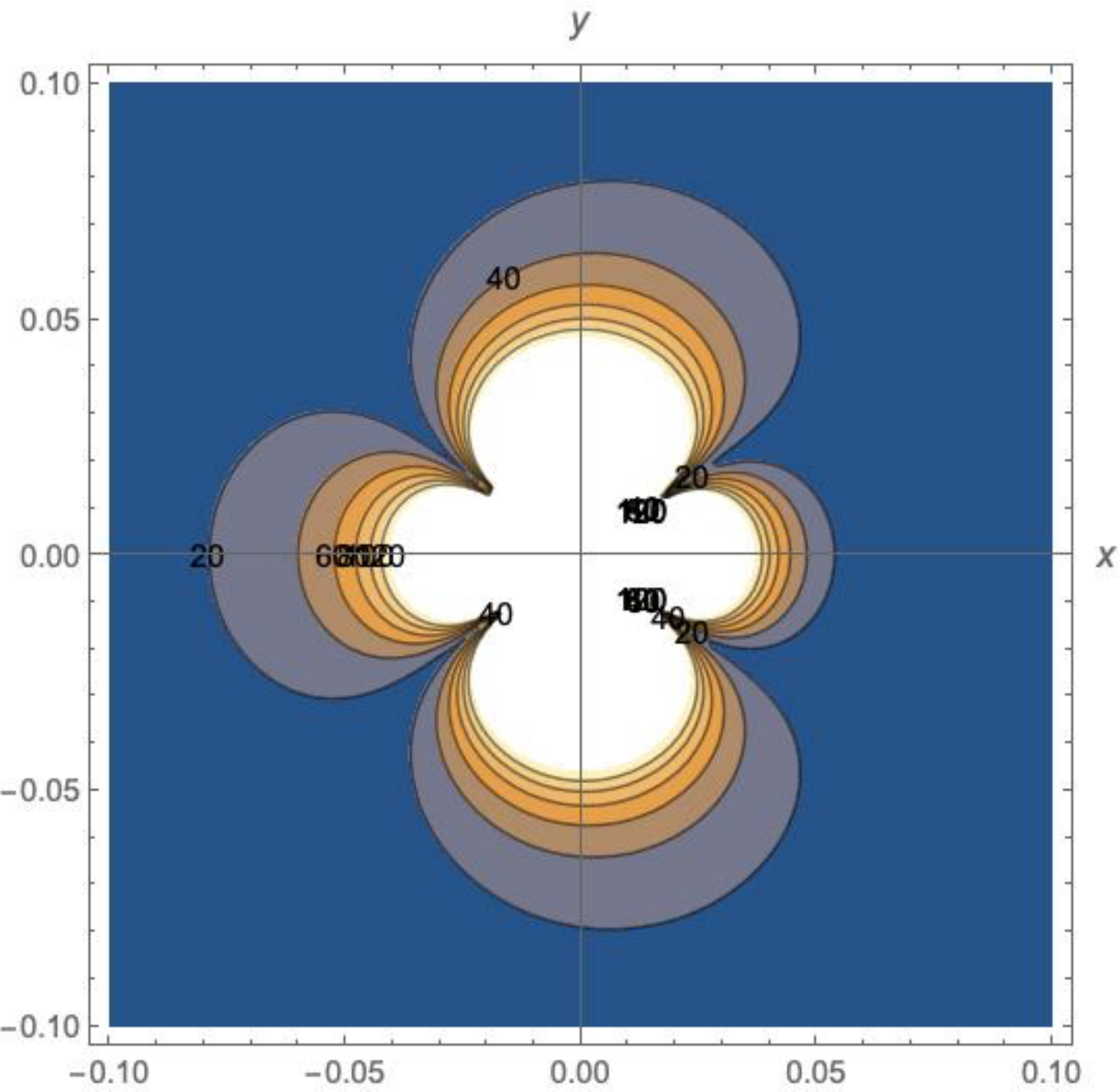}
\caption{(Color online)  Contours of constant normal current  {\it values} $\sqrt{g_{nx}^2 +g_{ny}^2 }$ for a  moving vortex with $s=0.05$ at short distances. $x$ and $y$ are in units of $\Lambda$.}
\label{f7}
\end{figure}

Fig.\,\ref{f5} show  the stream lines of normal currents (in fact these are contours of constant $S(x,y)$) for a fast and slow motion.

  Figs.\,\ref{f6} and \ref{f7} show the distribution of normal current {\it values} at large and short distances for $s=0.05$. The exotic shape of these distributions   close to the core  signals to possible peculiarities of supercurrents as well.

\begin{figure}[h ]
\includegraphics[width=7cm] {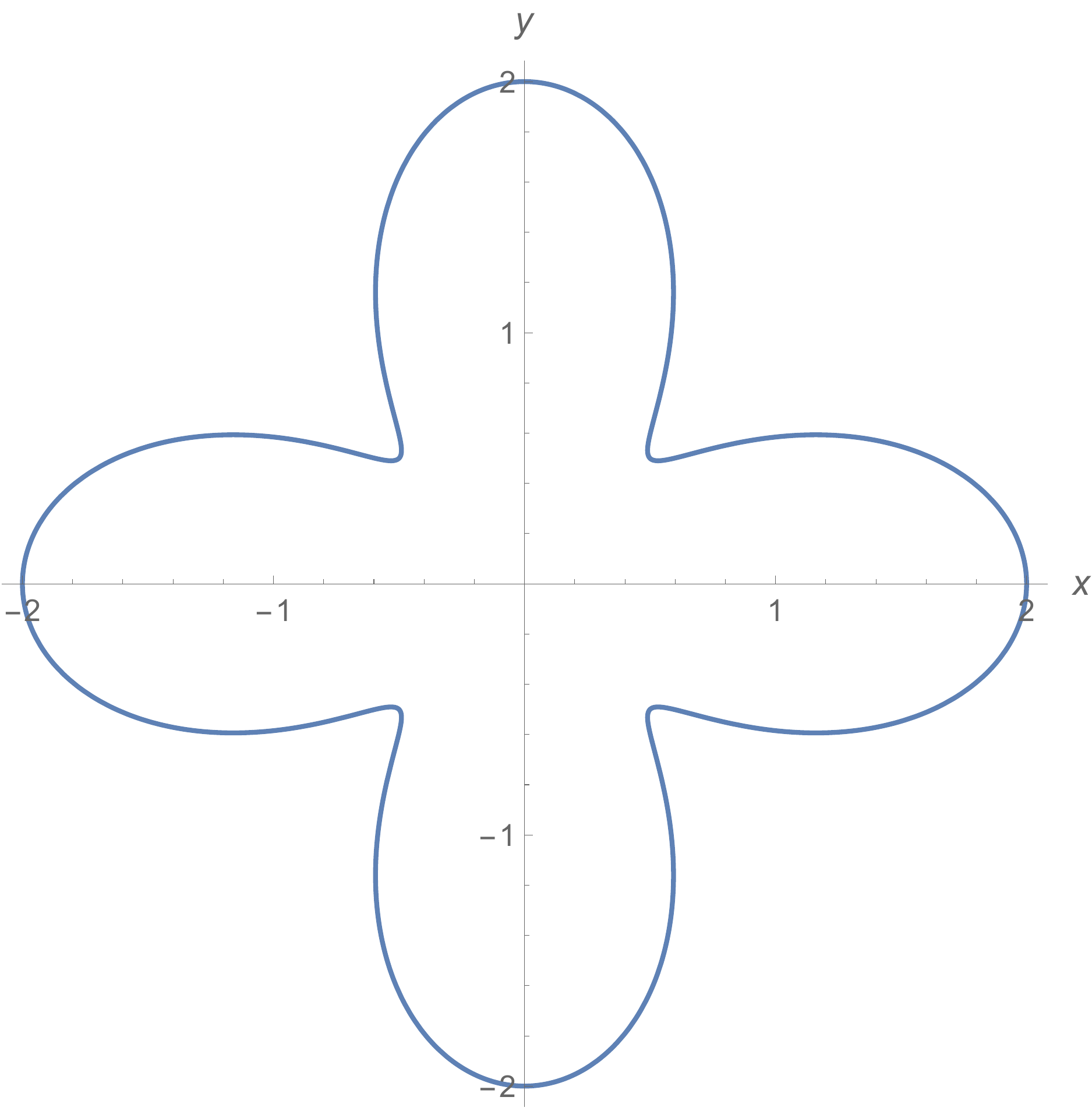}
\caption{Contours of constant  {\it values} $\sqrt{E_x^2 +E_y^2 }$ according to Eq.\,(\ref{Eyx1}) with $A=0.7$, $B=2$, and $C=4$.  }
\label{f8}
\end{figure}

 To make sense of the ``cloverleaf" shape of the normal current distributions, let us look at the $\bm E$ components of Eqs.\,(\ref{Ex}), (\ref{Ey}) for slow motion $s\ll 1$. Note that the pre-factor $\phi_0v/2\pi \lambda^2c \propto v \propto s$, so that for $s\to 0$ one can set  $s=0$ in  the integrals over $u$:
\begin{eqnarray}
&&  E_x\propto -s  \int_0^\infty du\,e^{-u}  \frac{y x   (u+2{\cal R}_1) }{
{\cal R}_1^3(u+{\cal R}_1)^2}, 
\label{Ex2}\\
&& E_y 
 = s\int_0^\infty du\,e^{-u}   \frac{{\cal R}_1^2 (u+{\cal R}_1)- x ^2 (u+2{\cal R}_1)}{{\cal R}_1^3(u+{\cal R}_1)^2} .  \qquad  
\label{Ey2}
 \end{eqnarray}
 where ${\cal R}_1^2= r^2+u^2$. We note that in polar coordinates $x=r\cos\varphi,\,\,\,y=r\sin\varphi$, ${\cal R}_1$ depends only on $r$. Hence, we can write these equations as
 \begin{eqnarray}
   E_x = A(r)\sin 2\varphi, \quad E_y = B(r) - C(r)\cos^2\varphi\, \qquad  
\label{Eyx1}
 \end{eqnarray}
where $A(r),B(r),C(r)$ should be evaluated by integrations over $u$. But even without   integrations it is clear that the value $|\bm E|=\sqrt{E_x^2+E_y^2}$ depends on the azimuth $\varphi$. The same is true for $|\bm g_n|$. A simple example of the azimuthal dependence is shown in Fig.\,\ref{f8} for some set of $A,B,C$. 
Thus, the unusual distributions of $|\bm g_n|$ at short distances from the vortex center can be traced to azimuth dependent divergences of $\bm g_n$ when one approaches the vortex core.

\subsection{Persistent currents}

The normal sheet current density is $\bm g_n=\sigma \bm E\, d$, whereas the supercurrent is $\bm g_s= \bm g -\bm g_n$, so that
\begin{eqnarray}
 g_{sx}=  g_x -  g_{nx} = \frac{c\phi_0 }{\Lambda^2}(\hat{g}_x-\frac{s}{4\pi^2}\hat{E}_x)\,, 
 \label{C18} 
 \end{eqnarray}
where $\hat{g}_x$ and $\hat{E}_x$ are the dimensionless integrals of Eqs.\,(\ref{gx1}) and (\ref{Ex}).  In particular, this reflects the fact that normal currents disappear at $v=0$.
Similarly, we have
\begin{eqnarray}
 g_{sy}=  g_y -  g_{ny} = \frac{c\phi_0 }{\Lambda^2}(\hat{g}_y-\frac{s}{4\pi^2}\hat{E}_y)\,.
 \label{C18a} 
 \end{eqnarray}

\begin{figure}[h ]
\includegraphics[width=7.5cm] {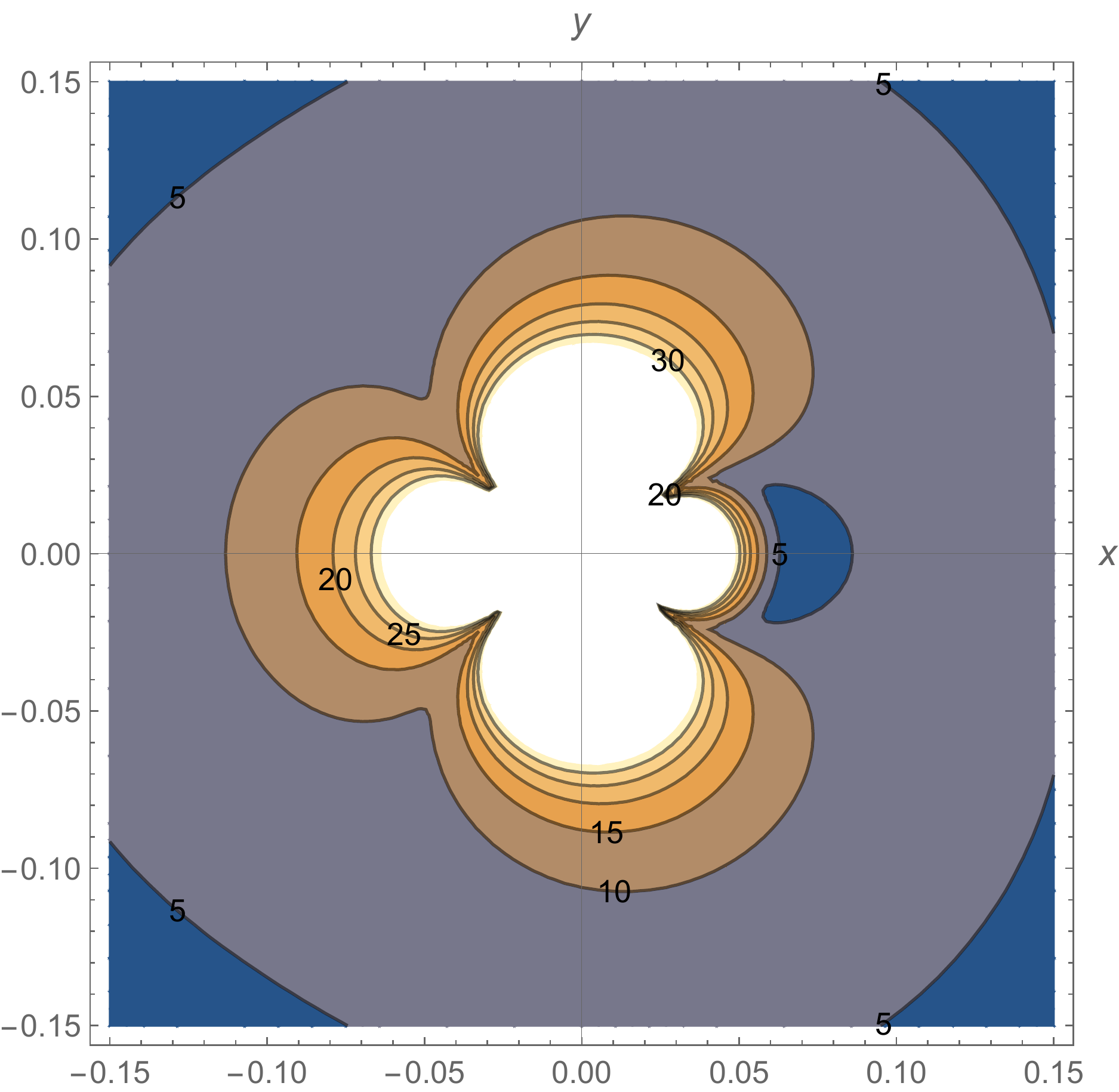}
\caption{ (Color online)  Contours of constant supercurrent  {\it values} $\sqrt{g_{sx}^2 +g_{sy}^2 }$ for a  moving vortex with $s=0.05$.  $x$ and $y$ are in units of $\Lambda$. One can see that the anisotropy of this distribution decreases with the distance from the vortex singularity.}
\label{f9}
\end{figure}

\begin{figure}[h ]
\includegraphics[width=7.5cm] {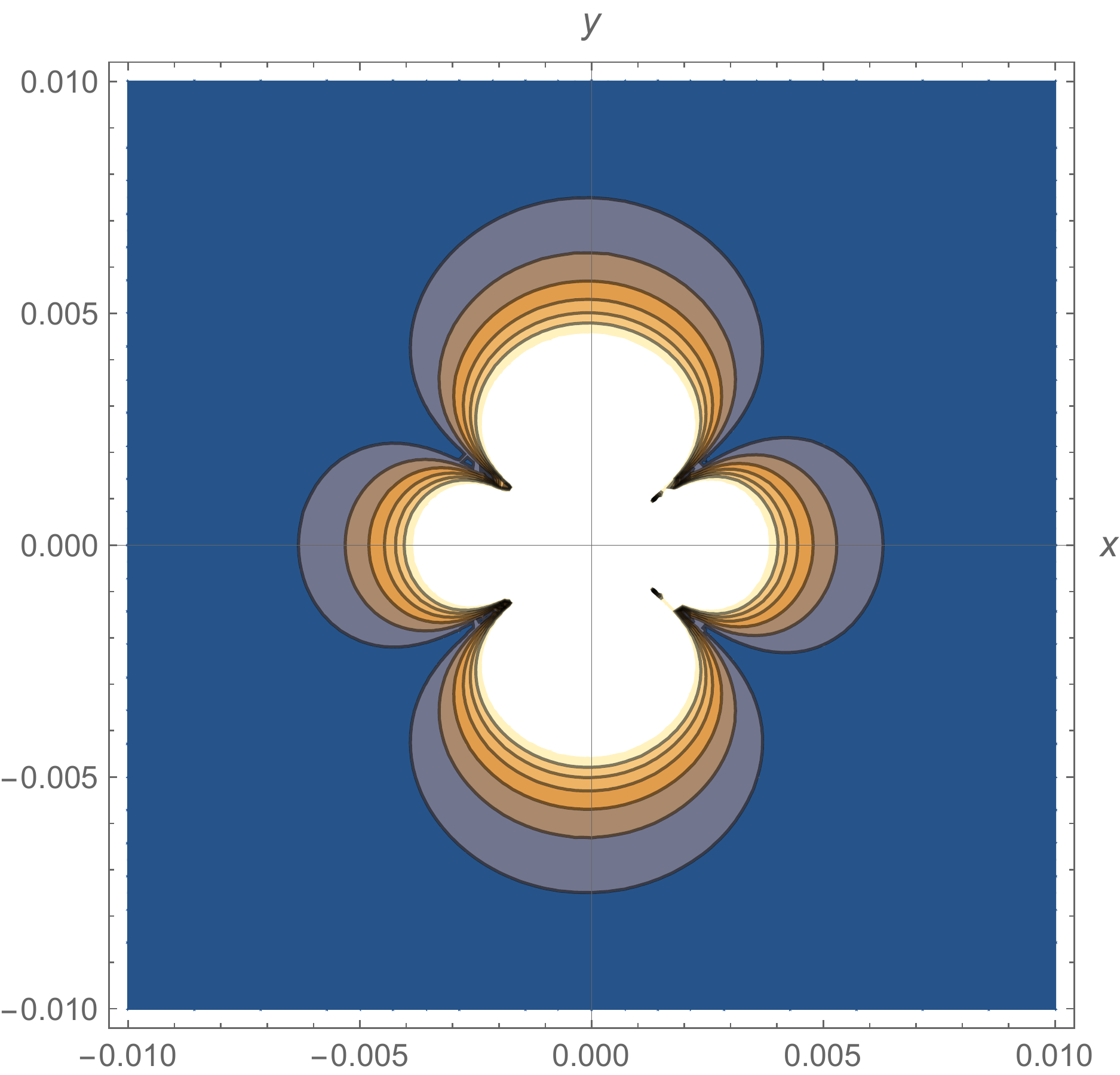}
\caption{(Color online)   Contours of constant supercurrent  {\it values} $\sqrt{g_{sx}^2 +g_{sy}^2 }$ for a  moving vortex with $s=0.05$ at short distances.  $x$ and $y$ are in units of $\Lambda$. }
\label{f10}
\end{figure}

As mentioned, the vortex core cannot be described by the London theory, but the distribution of $\bm g_s (\bm r)$ outside the core  may affect its shape \cite{Ichioka}.
A qualitative picture of the core {\it shape} can be obtained by examining contours of the current {\it values} $|\bm g_s(x,y )|=\,\,\,$const outside the core. This is shown in Fig.\,\ref{f9} and \ref{f10}.

One can see that the distortion of supercurrents  near the vortex  core disappears with increasing distance from the core.

The London current $g_s$ diverges if $r\to 0$. One can define  qualitatively the core ``boundary" as a curve where the London current reaches the depairing value, $j_d= c\phi_0/16\pi^2 \lambda^2\xi$ for the bulk or $g_d= c\phi_0/8\pi^2 \Lambda\xi$ for thin films, see e.g. Ref.\,\cite{Ichioka}. For the Abrikosov vortex at rest in isotropic  bulk case this simple procedure gives the coherence length $\xi$ as the core size. 
An example of the core shape so defined for a moving vortex is given in Fig.\,\ref{f10}. 

The persistent currents of the Pearl vortex at rest diverge   as $const/r$ when $r\to 0$, see e.g. \cite{jjf}. But for a moving vortex, the constant here may depend on the direction along which the origin is approached. As we have seen, the total magnetic flux carried by vortex is $\phi_0$, therefore, appearance of normal currents in moving vortex will cause redistribution of supercurrents as well, in particular, near the singularity this redistribution is substantial because normal currents diverge there, too, see Figs.\,\ref{f6} and \ref{f7}. This may result in   
  exotic shapes of the lines of constant $|g_s(x,y)|$ at short distances from the singularity.    

\section{Discussion}

We have shown that  the magnetic structure of the moving  vortices   is distorted relative to the vortex at rest.   The flux quantum of a moving  vortex is redistributed, the back side part of the flux is enhanced, whereas the in-front part is depleted. Physically, the distortion is caused by normal currents arising due to  
changing in time magnetic field at each  point of space, the electric field is induced and causes normal currents. 
 
Distributions of both  normal and persistent currents have  been considered for moving Abrikosov vortices in the bulk and for Pearl vortices in thin films. It turned out for films that these distributions at distances $r$ small on the scale of the Pearl $\Lambda$ have exotic shapes shown in Figs.\,\ref{f8}, \ref{f9}. 

This finding is potentially relevant because the vortex core {\it shape} might be  affected by persistent currents out of the core, if the core ``boundary" is defined as the place where the outside supercurrents reach the depairing value, the concept introduced by L. D.  Landau in his theory of superfluidity \cite{LL}. Our calculations show that if one approaches the vortex singularity along a straight line at an angle $\varphi$ with velocity (directed along $x$), the depairing value is reached at the azimuth dependent $r(\phi)$. Validity of such a definition of the core boundary should be confirmed, of course, by microscopic calculations of the order parameter $\Delta(r,\varphi)$ {\it inside} the core, the problem out of the scope of this work. Our intention is to address this question in near future. 
  
  If our picture is confirmed, the problem will arise about the structure of the   inside-the-core quasi-particle states in moving vortices which can differ substantially  from the states in the vortex at rest \cite{deGennes1}. 
 
Uncommon core current distributions of moving vortices in isotropic materials is due to the vector of velocity that breaks the cylindrical symmetry of fields and currents. In other words, the problem  becomes anisotropic. This leads to an idea that in {\it anisotropic} superconductors similar distributions   might occur even in vortices at rest. We plan to present our results for this case in a separate publication.

There are  experimental techniques which, in principle, could probe the field distribution in moving vortices \cite{Eli}. This is a highly sensitive SQUID-on-tip  with the  loop small on the scale of possible Pearl lengths.  
Recent experiments have traced  vortices moving in thin   films with   velocities well above the speed of sound \cite{Eli,Denis}. Vortices crossing thin-film bridges being pushed by transport currents   have a tendency to form chains directed along the velocity. The spacing of vortices in a chain is usually exceeds by much the core size, so that commonly accepted reason for the chain formation, namely, the  depletion of the order parameter behind moving vortices    is questionable. But at distances $r\gg \xi$ the time dependent London theory is applicable. Another promising technique of studying moving vortices could be Tonomura's Lorentz Microscopy \cite{Tonomura}.

 \section{Acknowledgements}
The work of V.K. was supported by the U.S. Department of Energy (DOE), Office of Science, Basic Energy Sciences, Materials Science and Engineering Division.  Ames Laboratory  is operated for the U.S. DOE by Iowa State University under contract \# DE-AC02-07CH11358.



\begin{thebibliography}{99}
\bibitem{remark} Throughout this paper we use numerical integrations for integrals over $0<u<\infty$ similar to that in the first line of Eq.\,(\ref{hz(r)1b}) to obtain fields or current distributions singular at the origin. Since this equation also provides the exact result in the second line, we compared  analytic outcomes  with numerical ones obtained with the help of Mathematica and found relative  differences on the order of $10^{-10}$.
\bibitem{remark2} The electric field in vacuum can be neglected since out of the film $E\sim vH/c$ and $v\ll c$. In other words, the electro-magnetic radiation is disregarded.
\end{thebibliography}
\end{document}